\documentclass[%
 aip,
 amsmath,amssymb,
 reprint,%
]{revtex4-1}

\usepackage{graphicx}
\usepackage{dcolumn}
\usepackage{bm}

\usepackage{mathptmx}
\usepackage{etoolbox}

\usepackage{xcolor}

\makeatletter
\def\@email#1#2{%
 \endgroup
 \patchcmd{\titleblock@produce}
  {\frontmatter@RRAPformat}
  {\frontmatter@RRAPformat{\produce@RRAP{*#1\href{mailto:#2}{#2}}}\frontmatter@RRAPformat}
  {}{}
}%
\makeatother
\begin{document}

\preprint{AIP/123-QED}


\title[Human behavior-driven epidemic surveillance in urban landscapes]{Human behavior-driven epidemic surveillance in urban landscapes}

\author{P. Valga\~n\'on}%
\affiliation{Department of Condensed Matter Physics, University of Zaragoza, 50009 Zaragoza (Spain).}

\affiliation{GOTHAM lab, Institute for Biocomputation and Physics of Complex Systems (BIFI), University of Zaragoza, 50018 Zaragoza (Spain)}

\author{A.F. Useche}
\affiliation{Department of Health Management and Policy, Dornsife School of Public Health, Drexel University,
Philadelphia, PA 19104 (USA)}

\affiliation{Department of Industrial Engineering, School of Engineering, Universidad de Los Andes, 111711 Bogot\'a
(Colombia)}

\author{F. Montes}
\affiliation{Department of Industrial Engineering, School of Engineering, Universidad de Los Andes, 111711 Bogot\'a
(Colombia)}

\author{A. Arenas}
\email{alexandre.arenas@urv.cat}
\affiliation{Departament d'Enginyeria Inform\`atica i Matem\`atiques, Universitat Rovira i Virgili, 43007 Tarragona (Spain)}

\author{D. Soriano-Pa\~nos}
\email{sorianopanos@gmail.com}
\affiliation{Departament d'Enginyeria Inform\`atica i Matem\`atiques, Universitat Rovira i Virgili, 43007 Tarragona (Spain)}
\affiliation{GOTHAM lab, Institute for Biocomputation and Physics of Complex Systems (BIFI), University of Zaragoza, 50018 Zaragoza (Spain)}
\affiliation{Instituto Gulbenkian de Ciencia, 2780-156 Oeiras (Portugal)}

\author{J. G\'omez-Garde\~nes}
\email{gardenes@unizar.es}
\affiliation{Department of Condensed Matter Physics, University of Zaragoza, 50009 Zaragoza (Spain).}
\affiliation{GOTHAM lab, Institute for Biocomputation and Physics of Complex Systems (BIFI), University of Zaragoza, 50018 Zaragoza (Spain)}
\affiliation{Center for Computational Social Science, University of Kobe, 657-8501 Kobe (Japan)}

\date{\today}

\begin{abstract}
We introduce a surveillance strategy specifically designed for urban areas to enhance preparedness and response to disease outbreaks by leveraging the unique characteristics of human behavior within urban contexts. By integrating data on individual residences and travel patterns, we construct a Mixing matrix that facilitates the identification of critical pathways that ease pathogen transmission across urban landscapes enabling targeted testing strategies. Our approach not only enhances public health systems’ ability to provide early epidemiological alerts but also underscores the variability in strategy effectiveness based on urban layout. We prove the feasibility of our mobility-informed policies by mapping essential mobility flows to major transit stations, showing that few resources focused on specific stations yields a more effective surveillance than non-targeted approaches. This study emphasizes the critical role of integrating human behavioral patterns into epidemic management strategies to improve the preparedness and resilience of major cities against future outbreaks.
\end{abstract}

\maketitle



Throughout history, the interplay between epidemics and human societies has been profound, each significantly influencing and shaping the course of the other \cite{Diamond1997,Snowden2019,McNeill1976}.
This interplay, already present during our early hunter-gatherer days, was notably boosted by the establishment of agrarian societies around 10,000 years ago, when the creation of communities provided fertile breeding grounds for diseases to thrive. It was during this transformative period that humanity first encountered diseases such as malaria, tuberculosis, leprosy, influenza, and smallpox, which have since significantly altered the course of human history \cite{DominguezAndres2021}. 

As human settlements transformed into nowadays vast urban centers, they also introduced new challenges for contemporary epidemiology \cite{Alirol2011,Lee2020}. Urban environments, characterized by dense populations, complex social interactions, and socioeconomic disparities, create ideal conditions for the spread of communicable diseases \cite{Brizuela2021,Bilal2021,Kache2022}. Moreover, the rapid movement of people and goods within and across cities sweeps out national and continental boundaries, thus facilitating the global dissemination of pathogens between major city centers \cite{Baker2022}. 

The advent of big data and advanced modeling techniques offers new avenues for understanding and managing this cocktail of epidemic boosters, emphasizing the importance of integrating mobility data and demographic insights into urban epidemic management strategies \cite{buckee2021thinking}. In this line, data analytics have significantly advanced epidemic modeling \cite{surveillance,prudent}, enabling the mathematical formalization of complexities associated with demographic segregation, mobility patterns, and heterogeneous contacts—key factors in disease propagation~\cite{ManfrediDOnofrio2015}.  

\begin{figure*}[t!]
\centering
\includegraphics[width=1.00\linewidth]{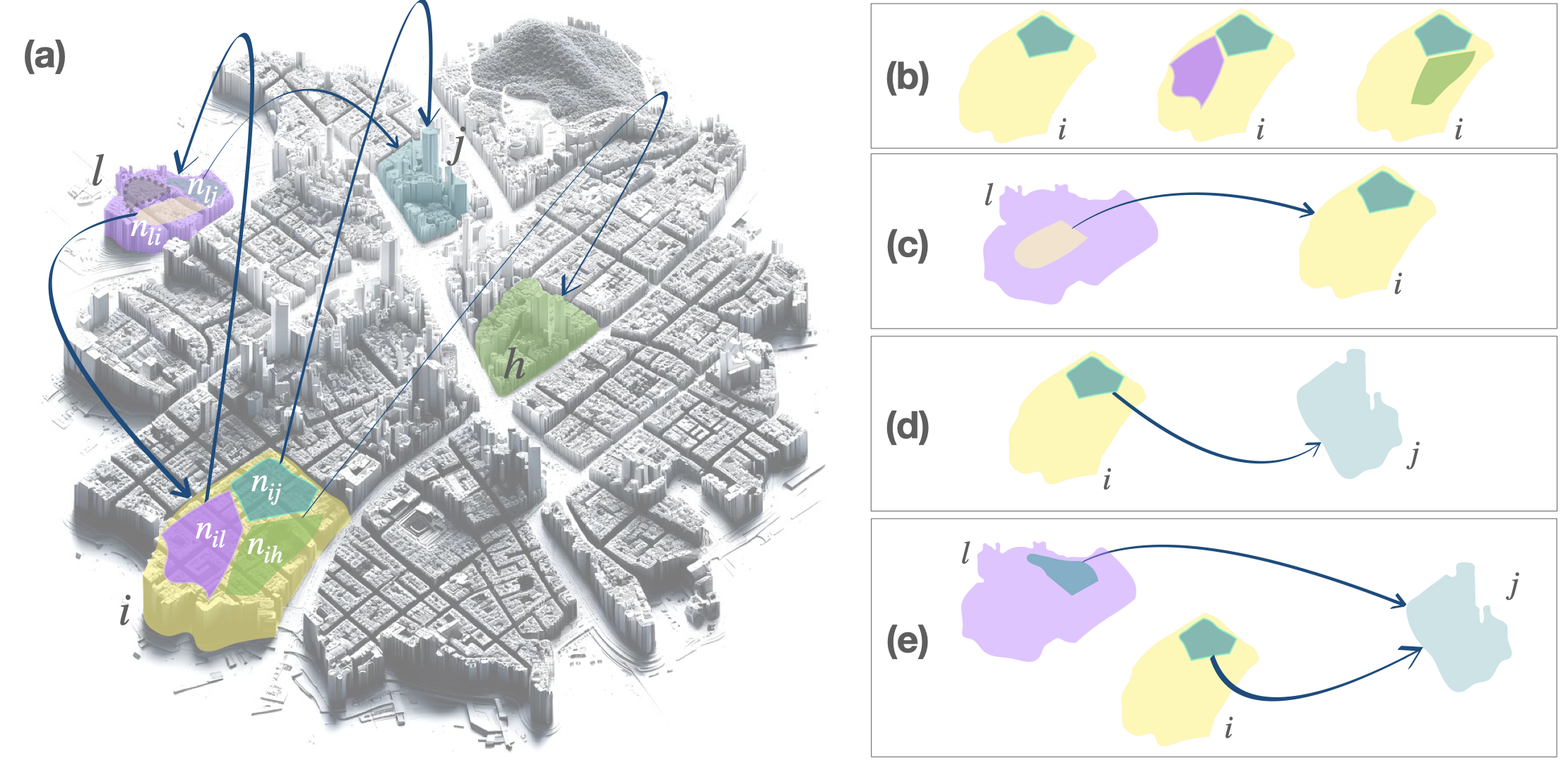}
\caption{{\bf Mobility model and mixing of the population.} (a): Illustration of the mobility model here considered. We assume a metapopulation of $N$ patches and construct the mobility network from daily back-and-forth movements recorded in public surveys. Each geographical patch $i$ in the metapopulation is partitioned into different subgroups according to their usual destination, being $n_{ij}$ the number of residents inside $i$ daily commuting to patch $j$. (b)-(e): Mixing patterns of the subpopulation of $n_{ij}$ residents in $i$ with typical destination in $j$. When staying in their residential patch $i$, these residents might interact with others staying also there (b) and also with residents from other patches (c) that have as usual destination patch $i$. In turn, when moving, residents in $i$ will interact with those residents from other patches: (d) either those that belong to the usual destination $j$ and that decide not to move or (e) with residents of a different patch $l\neq j$ who also have as usual destination patch $j$.}
\label{fig:1}
\end{figure*}

Utilizing metapopulation frameworks, epidemic models effectively manage the interaction between the above ingredients by coupling contact-driven transmission with mobility-related dispersal \cite{Watts2005,Colizza2007,ColizzaVespignani2007,Colizza2008,Balcan2009,BelikGeisel2011,Meloni2011,Castioni2021CriticalBehavior}. In recent decades, such behaviorally informed metapopulation models have become indispensable for mechanistic forecasting, enabling precise predictions of epidemic trajectories across diverse scales—from local communities \cite{SorianoPanos2018,SorianoPanos2020} through national \cite{Gatto2020,Bertuzzo2020,Arenas2020} to global levels \cite{Colizza2006,Brockmann2013,Zhang2017SpreadZika}. 
Moreover, contemporary epidemic models extend beyond mapping the spatio-temporal spread of diseases, but also facilitate the development of data-informed containment strategies with maximal resource efficiency and minimal socioeconomic disruption~\cite{Bosetti2020,zhu2021allocating,ReynaLara2022,Mazzoli2023}.

These efforts highlight the critical role that incorporating social dynamics into epidemic models has on our capacity to respond to infectious disease threats, an aspect that has recently materialized in research agendas \cite{agenda,outlook,salathe2018digital} aimed at advancing this endeavor. These agendas, among other problems, specifically highlight the challenge of employing realistic human contact structures to explore localization behavior in {\em key subpopulations} for epidemic control policies. Addressing this precise challenge, here we introduce a data-informed approach tailored for large urban centers, leveraging their unique demographic and mobility patterns. Our framework allows identifying critical human flows and leads to actionable surveillance strategies to obtain valuable early warnings of incoming epidemic outbreaks.

\section*{Formalization}
\label{sec:formalization}

To address the former challenge, we first present a formalism aimed at capturing the complex spatiotemporal structure of human mobility, particularly within urban environments. Over the last decades, extensive research has been devoted to reveal the complex and rich spatiotemporal structure of human mobility patterns \cite{Gonzalez2008UnderstandingMobility,Barbosa2018HumanMobility}. When focusing on urban environments, public surveys reveal that human flows are mainly dominated by recurrent mobility patterns \cite{Jiang2016TimeGeo,Bokanyi2021UniversalPatterns} connecting the residence of individuals with the usual work place to which they commute. Consequently, recurrent human flows have been incorporated into theoretical frameworks \cite{Balcan2011PhaseTransitions,Belik2011,ApolloniPoletto2014,Charaudeau2014CommuterMobility,GomezGardenes2018CriticalRegimes} as the backbone of metapopulation models tailored to track urban epidemics. This way, the analysis of origin-destination matrices (OD), either constructed from mobile phone devices or from census surveys, have been instrumental in shedding light into the vulnerability of cities during the initial stages of epidemics and the spatially uneven distribution of cases  \cite{scarpinoCOVID,Hazarie2021InterplayDensityMobility,Aguilar2022UrbanStructure}. 

Regardless of the former advances, the feasibility and efficiency of surveillance campaigns and control strategies driven by commuting data remains to be explored. To tackle this challenge here we leverage the formalism proposed in \cite{Valganon2022ContagionDiffusion} for the study of contagion-diffusion processes with recurrent mobility patterns of distinguishable agents. This model represents a city as a collection of $P$ interconnected patches (see Fig.~\ref{fig:1}a), each one representing a geographical area. The mobility data informs the matrix {\bf n}, whose elements, $n_{ij}$, identify subgroups of individuals living inside patch $i$ and regularly travelling to patch $j$. Therefore, a generic patch, say $i$, is populated by $N_i$ residents, divided into subgroups based on their travel destinations ($N_i=\sum_j n_{ij}$), as schematized in Fig.~\ref{fig:1}.a.

\begin{figure*}[t]
    \centering
    \includegraphics[width=0.99\linewidth]{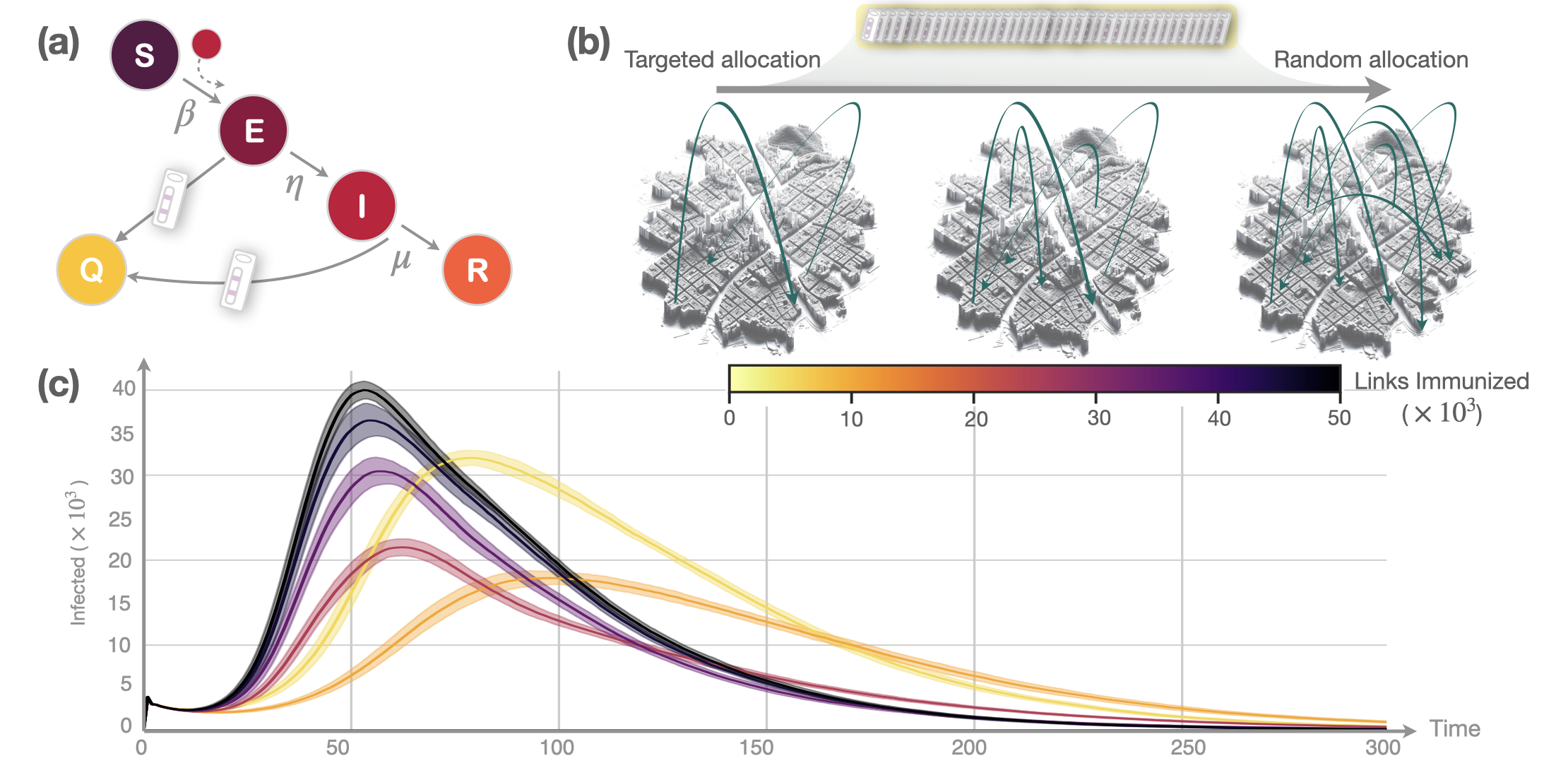}
    \caption{{\bf Mobility-informed testing and quarantine policies}. (a) Susceptible-Exposed-Infected-Recovered-Quarantined (SEIRQ) compartmental model used in the epidemic simulations. In addition to the usual flows governed by the pathogen infectiousness $\lambda$, the incubation period $\eta^{-1}$ and the recovery probability $\mu$, we assume that Infected and Exposed individuals are quarantined after being tested. (b) Illustration of the testing strategy implemented in the model. Such strategy depends on the amount of test available to screen the population $n_{tests}$ and on how these test are allocated across the city, governed by $L$. We explore different scenarios, ranging from targeted resources allocation on the most vulnerable subgroups defined from the Mixing matrix ${\bf M}$  (small values of $L$) to random testing of the population (large $L$ values). (c) Time evolution of the number of infected individuals in Bogot\'a when distributing $n_{tests}=2\times 10^5$ tests on the $L$ most critical human flows (color code). Solid lines represent the average time evolution and the shadowed region corresponds to 95\% confidence interval over 25 simulations for each $L$ value. The basic reproduction number of the pathogen responsible for the outbreak is ${\mathcal R}_0=4$.} 
    \label{fig:tests}
\end{figure*}

Once constructed the matrix ${\bf n}$, we can simulate reactive processes, such as contagion dynamics, coupled with diffusion of the population. To this aim, the model assumes a discrete-time approach, where each time step consists of four different stages. First, each individual decides to move to the usual destination with probability $p_d$ or stay in the residence area. Second, (day) interactions take place with those agents sharing the same location. Third, those individuals that moved in the first stage return to their residence and, fourth, another interaction stage takes place when agents contact within their households (night interactions). In any of the two interaction stages occurring in each time step, contagions can occur. A more exhaustive description of the steps present in the simulation scheme is found in the Methods section.

The processes described above can be encapsulated into a Mixing matrix ${\bf M}$ which contains the expected number of interactions among the different subgroups of individuals found in an urban environment. We illustrate in Figs~\ref{fig:1}b-e the elements of the mixing matrix associated to an specific subgroup of individuals with residence in patch $i$ and destination in patch $j$. 
These elements capture all the contagion venues affecting these individuals as a result of their mobility across the city.
Namely, these individuals can interact with other residents staying there (Fig.~\ref{fig:1}.b) and with others coming from a different patch (Fig.~\ref{fig:1}.c). In contrast, when moving to their usual destination, they interact with its residents (Fig.~\ref{fig:1}.d) and with other visitors also moving there (Fig.~\ref{fig:1}.e). The mathematical expression of the elements of the Mixing matrix ${\bf M}$ can be found in the Methods section whereas their derivation from the equations governing the epidemiological model used here is explained in the Supplementary material.

\section*{Results}

\begin{figure*}[t]
    \centering
    \includegraphics[width=0.99\linewidth]{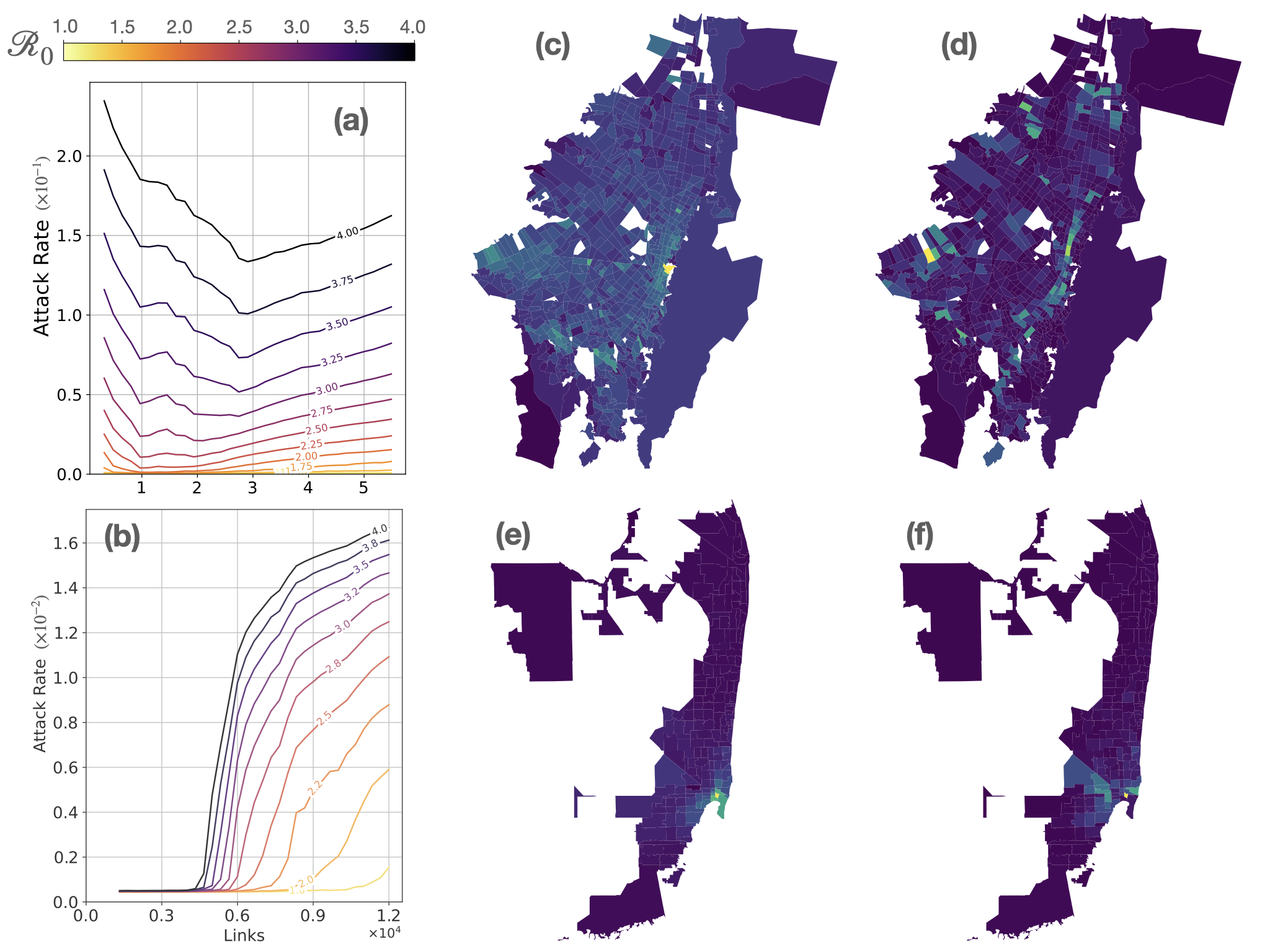}
    \caption{{\bf Urban landscapes and optimal distributions of tests across mobility flows} Panels (a) and (b) show the attack rate of an epidemic wave as a function of the basic reproduction number ${\cal{R}}_0$ of the pathogen (color code) and the number $L$ of mobility links used to distribute the $n_{tests}$ tests in Bogot\'a (a) and Miami (b). The central maps show the distribution of cases across Bogot\'a (c) and Miami (e) by plotting the attack rate in each patch after implementing a control strategy that sets $n_{tests}$ tests across the most important human flows ($L=2.8\times 10^4$ for Bogot\'a and $L=6\times 10^3$). Finally, the maps in the right show how the $n_{tests}$ were distributed according to the destination of the aforementioned most critical flows in Bogot\'a (d) and Miami (f). In all the maps, brighter colors correspond with higher values of the represented quantities.}
    \label{fig:attackrate}
\end{figure*}

\subsection{Mobility-informed testing and quarantine policies}

The leading eigenvector of Mixing matrices, derived from epidemic models on networked populations, plays a pivotal role in pinpointing key actors in epidemic outbreaks, whether they are individuals \cite{torres2021nonbacktracking} or specific geographical areas \cite{ReynaLara2022,zhu2021allocating}. Expanding upon these insights, we introduce a testing strategy that prioritizes screening of those agents participating in human flows with a large contribution to the leading eigenvector of the matrix ${\bf M}$. The operational framework of this policy is the Susceptible-Exposed-Infected-Recovered-Quarantine (SEIRQ) model, as depicted in Fig.~\ref{fig:tests}.a, so that those agents that test positive are isolated from the population as a control policy. 

As illustrated in Fig.~\ref{fig:tests}.b, our testing policy hinges on two critical parameters: the total number of tests $n_{\text{tests}}$ at our disposal, and the extent of human flows $L$ targeted by these tests. By adjusting the parameter $L$, our strategy can transition from a focused approach, targeting individuals taking part in those flows with the largest contributions to the leading eigenvector of ${\bf M}$, to a more distributed allocation of tests across the urban landscape. A more detailed exposition of the strategies for the distribution of testing resources is provided in the Methods section.

We apply our mobility-informed testing strategy to two real scenarios, the cities of Bogot\'a and Miami, whose mixing matrices are constructed from demographic and mobility data drawn from public surveys, as detailed in the Methods section. For this purpose, we simulate several outbreaks caused by a pathogen with basic reproduction number ${\cal R}_0=4$ and for which we have $n_{tests}=2\times 10^5$ tests to counteract its spread. This way, varying the number $L$ allows us to evaluate the impact of different testing distributions.

In Fig.~\ref{fig:tests}.c we plot the time evolution of the number of infected for different strategies. Starting from an scenario when resources are distributed quite evenly across the city ($L=5\times 10^4$), decreasing $L$ yields the usual flattening of the epidemic curves, reducing and delaying the peak of contagions. However, when tests are concentrated in a few critical flows ($L=2\times 10^3$), spreading them out by increasing $L$ also favors epidemic control. Overall, our analysis for the city of Bogot\'a reveals the existence of an optimal number of human flows $L^*_{\textrm{opt}}$ on which testing and isolation policies should be prioritized to efficiently control an outbreak. Remarkably, Figure S1 in the Supplementary Material shows that such $L^*_{\textrm{opt}}$ value does not exist for Miami, pinpointing that in this city the concentration of tests on a few subgroups of flows, i.e. reducing $L$, seems to be the best control strategy.

\subsection{Optimal control policies depend on urban landscapes}

To further showcase the effects of targeted testing and quarantine policies, we now focus on the impact of the mobility-informed policies on the attack rate, defined as the total number of individuals contracting the disease during an epidemic outbreak. Figures~\ref{fig:attackrate}.a-b show how the latter indicator depends on both the basic reproduction number of the disease ${\mathcal R}_0$ and the spatial allocation of the $n_{\text{tests}}$ across $L$ mobility links in Bogot\'a and Miami respectively. Fig.~\ref{fig:attackrate}.a reaffirms the existence of an optimal value $L^*_{\textrm{opt}} ({\mathcal R}_0)$ of targeted links reducing the long-term impact of epidemics on Bogot\'a. Interestingly, such value is not independent of the epidemic process but strongly varies with the basic reproduction number ${\mathcal R}_0$, yielding a complex interplay between the extent of an outbreak and the optimal spatial allocation of resources to reduce its burden. Conversely, Fig.~\ref{fig:attackrate}.b shows that, in Miami, concentrating testing efforts on a few critical flows allows mitigating an outbreak even for relatively large ${\mathcal R}_0$ values.

Our findings underscore that mobility-informed policies should be adapted to the unique characteristics of each urban environment \cite{Roth2011Structure,Bassolas2019Hierarchical}. Specifically, in cities with complex and segregated socio-economic activities, concentrating resources on the most critical flows does not always ensure community-wide protection and thus proves ineffective. Conversely, this strategy yields significant benefits in cities where the most relevant flows are primarily confluent towards a few and close patches. These two contrasting patterns correspond to the two cities under study and can be further elucidated by examining the spatial distribution of cases for a resource allocation characterized by $L\gtrsim L^{\star}_{\textrm{opt}} ({\mathcal R}_0)$. 

In Figure~\ref{fig:attackrate}.c, we depict the attack rate in each patch~$r_i^\infty$ in Bogot\'a, assuming that tests are allocated across the $L=2.8 \times 10^4$ most critical flows. This reveals a widespread penetration of the disease with varying degrees of impact across different areas. We also represent in Fig.~\ref{fig:attackrate}.d the spatial allocation of resources, that is, the proportion of tests conducted on individuals moving to each patch in this scenario, indicating that the mobility-informed policies prioritize acting over human flows spanning multiple neighborhoods. An in-depth exploration of the dependence of both cases and test distribution on $L$ is presented in Figure S2 in the Supplementary Material. There, it becomes evident that the optimal $L^{\star}_{\textrm{opt}}$ arises from the trade-off between the spatial allocation of resources required to address different contagion sources and the minimum number of tests needed for local outbreak control in each patch. In fact, when $L<L^*_{\textrm{opt}} ({\mathcal R}_0)$, the epidemic is contained in very specific locations but emerges elsewhere. Conversely, beyond the optimal value, i.e. when $L>L^*_{\textrm{opt}} ({\mathcal R}_0)$, the dispersion of tests across the urban landscape is such that preventing the spread in the main outbreak becomes totally unfeasible.

The epidemic scenario observed in Bogotá starkly contrasts with that of Miami for {$L=6\times 10^3$}, where the majority of both cases (Fig.~\ref{fig:attackrate}.e) and the allocated tests (Fig.~\ref{fig:attackrate}.f) concentrate around a single patch and its immediate vicinity. In this context, the aforementioned trade-off leans towards maximizing local outbreak control, thereby favoring strategies that adopt lower $L$ values. Specifically, Figure S2 reveals that as tests become more evenly distributed (i.e., as $L$ increases), the surge in attack rate shown in Figure~\ref{fig:attackrate}.b is attributable to a significant outbreak in the most vulnerable area, which cannot be contained with the limited resources available there.

\begin{figure*}[t]
    \centering
    \includegraphics[width=0.975\linewidth]{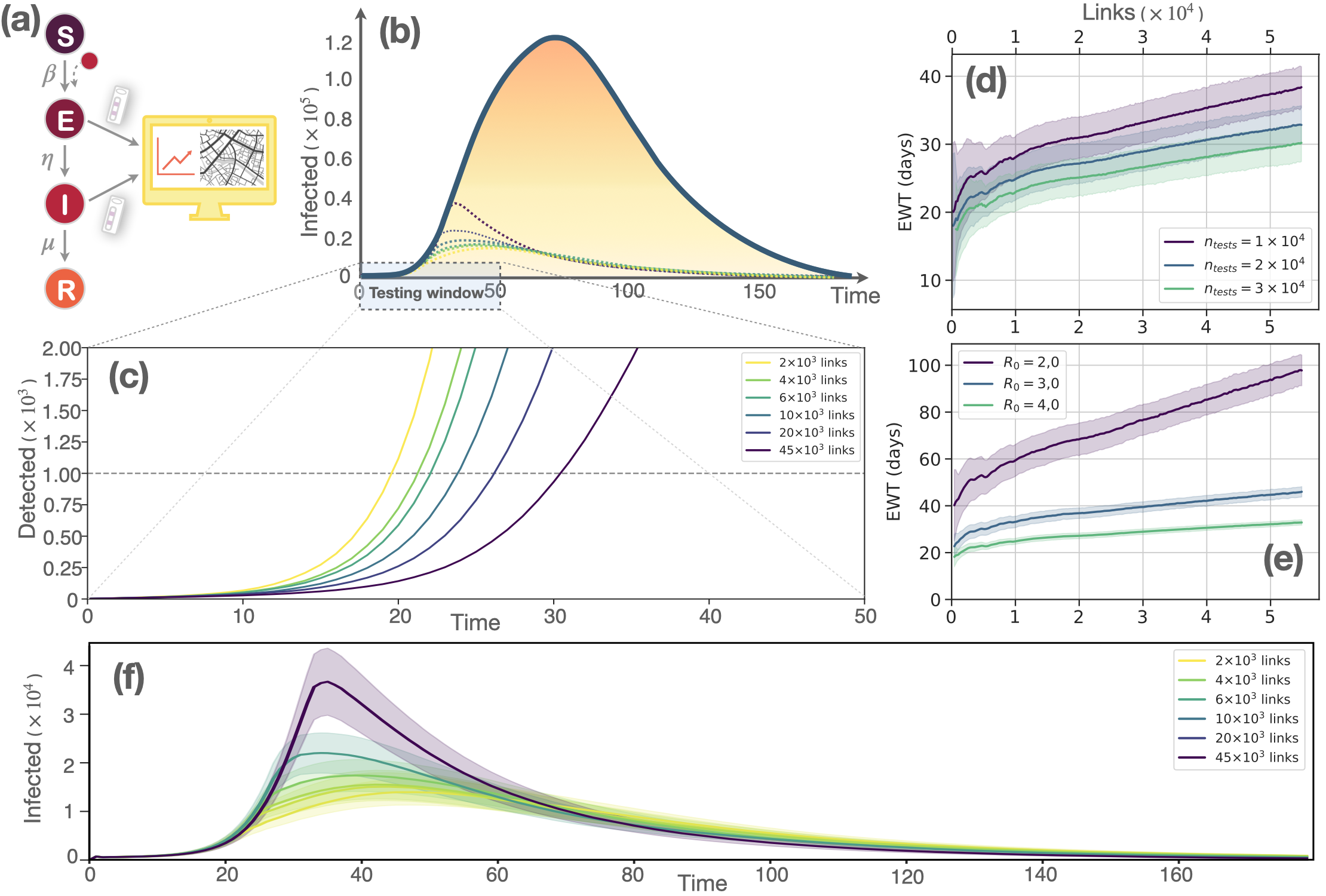}
    \caption{\textbf{Enhancing Epidemic Response with Targeted Surveillance.} (a) The SEIR model is augmented with a surveillance system that records positive tests from exposed and infectious individuals. (b) The solid line represents the natural (without interventions) epidemic curve, at the beginning of which (see window) tests are carried out. The dashed lines correspond to the epidemic curves shown in (f) (commented below). In panel (c) we show the time evolution of detected cases for different concentrations of testing resources on critical links. The dashed horizontal line signifies the threshold, $n_{\text{alarm}}$, which triggers containment actions. The Early Warning Time (EWT) - the interval spanned from the initial infections to the time when $n_{\text{alarm}}$ detections occur - is plotted against the degree of test concentration $L$ for (d) various numbers of available tests, $n_{\text{tests}}$, and for (e) distinct basic reproduction numbers, $\mathcal{R}_0$. Panel (f) shows the mitigated epidemic curves upon the imposition of containment measures that reduce ${\cal{R}}_0$ by $50\%$ at time $t=EWT$ (these curves are also represented with dashed lines in panel (b) for illustrating the mitigation effect with respect to the uncontrolled scenario). The shaded areas in panels (d-f) represent the 95\% confidence interval in the results obtained from 100 distinct simulations.}
    \label{fig:EWT}
\end{figure*}

\subsection{Mobility-informed surveillance policies}

The results shown above have illustrated that the Mixing matrix ${\bf M}$ can guide the optimal allocation of tests to reduce the long-term extent of epidemic outbreaks. Regardless of the potential adequacy of the different control policies here proposed, their timing of implementation also represents a crucial factor shaping their ultimate benefits to reduce the epidemic burden on the population~\cite{oraby2021modeling,steinegger2023joint,morris2021optimal}. Such timing is closely related to the design of efficient surveillance strategies, which should provide the authorities with an up-to-date picture of the evolution of the number of cases for the early diagnosis of ongoing epidemic crises. 

We now explore whether the Mixing matrix ${\bf M}$ can inform surveillance campaigns for an efficient screening of the infected population in an epidemic outbreak. In what follows, we restrict our analysis to Bogot\'a and consider an epidemic outbreak with $\mathcal{R}_0=4$. Our surveillance strategy distributes $n_{tests}=2\times 10^4$ tests across the $L$ most critical flows identified in the matrix {\bf M}. Unlike the previous analysis, we assume an uncontrolled scenario, meaning that infected individuals are no longer quarantined upon detection but, instead, their number is monitored for surveillance purposes as shown in the compartmental diagram in Fig~\ref{fig:EWT}.a. Thus, our baseline scenario (see Fig.\ref{fig:EWT}.b) consists in an uncontrolled epidemic wave in which contagions naturally propagate through the urban environment. This enables us to quantify the temporal evolution of positive cases detected during a {\em Testing window} (highlighted in Fig.\ref{fig:EWT}.b) as the epidemic unfolds without case detection having any effect on its evolution. 

In Fig.~\ref{fig:EWT}.c, we present the temporal evolution of the total number of detected cases for different allocation strategies of testing resources, regulated by $L$. These results indicate that concentrating testing resources on a limited number of key flows (indicated by a small $L$) enables surveillance systems to detect an impending epidemic more swiftly. In addition, the cumulative number of cases detected 30 days after the beginning of the outbreak decreases as tests are more evenly distributed across the city, for different values of $\mathcal{R}_0$ and amounts of testing  resources, $n_{\text{tests}}$ (see Figure S3 in the Supplementary Material).

To more precisely measure this effect, we introduce the concept of the {\em Early Warning Time} (EWT), defined as the time needed to declare an epidemic scenario after detecting $n_{\text{alarm}}$ infected individuals (indicated by the dashed line in Fig.~\ref{fig:EWT}.c). The increasing trend of the EWT with the number of links, $L$, across which tests are allocated is evident in Figs.~\ref{fig:EWT}.d and~\ref{fig:EWT}e. This observation aligns with the premise that mobility-informed strategies enhance the preemptive detection of an epidemic by targeting the testing of individuals in high-risk transmission flows. Moreover, these two panels corroborate the persistent benefit of such surveillance campaigns focused on the most critical human mobility flows, irrespective of the volume of testing resources, $n_{\text{tests}}$, (Fig.~\ref{fig:EWT}.d) and the basic reproduction number, $\mathcal{R}_0$, of the spreading pathogen (Fig.~\ref{fig:EWT}.e).

\begin{figure*}[t]
    \centering
    \includegraphics[width=0.99\linewidth]{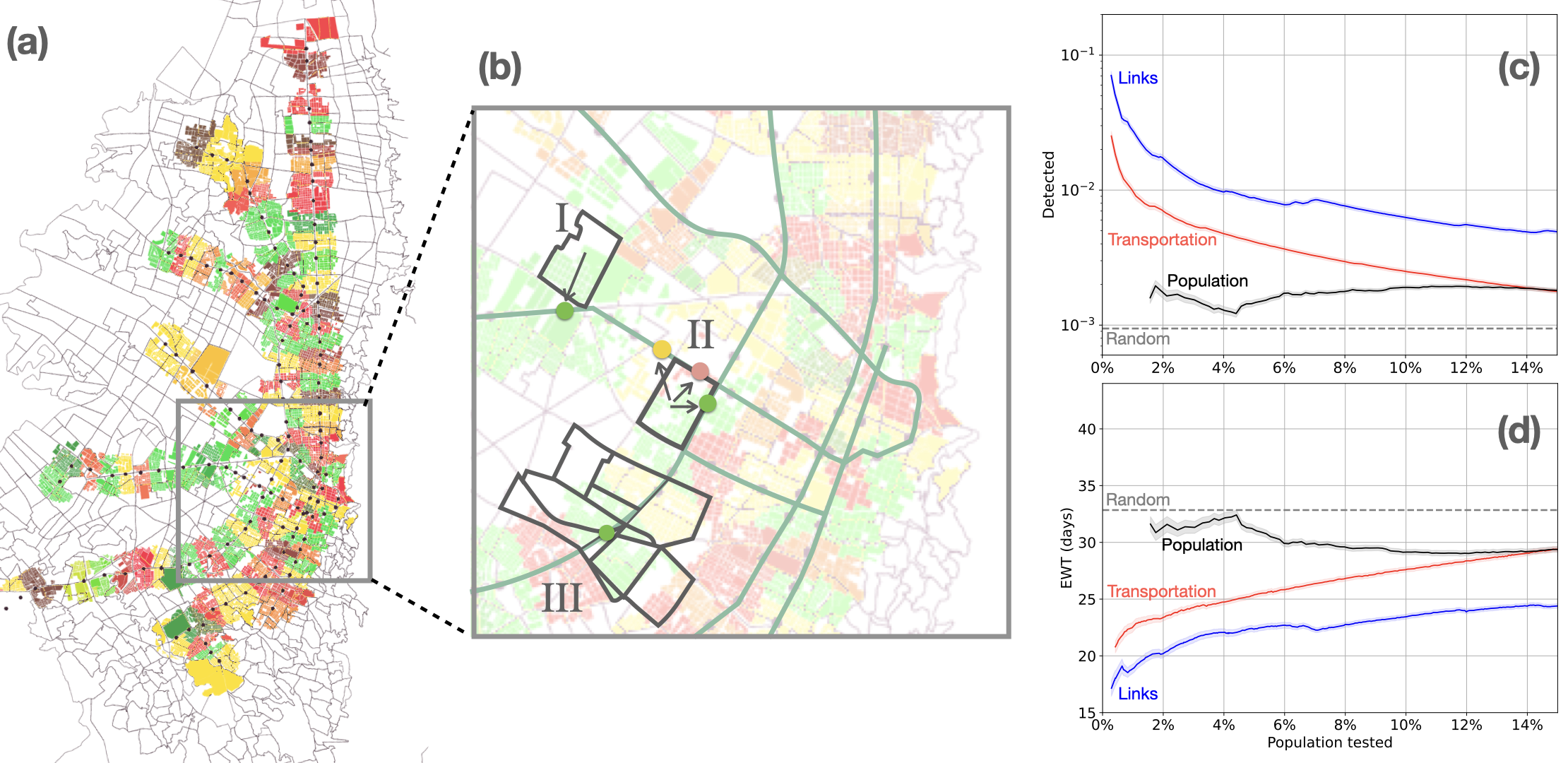}
    \caption{{\bf Transferring mobility-informed policies into massive transportation systems}. (a) The spatial distribution of the Transmilenio stations in Bogotá (Colombia): colored areas correspond to the different patches whereas dots correspond to stations. The alignment of raw mobility flows with Transmilenio journeys is made by assigning (see Methods) the fraction of population of a patch making use of a particular station. Different situations of the assignment processes are highlighted in (b): (I) all the population of a patch is assigned to the same station, (II) three stations serve a single patch,    and (III) a single station serves the population of a collection of patches. Finally, panels in the right show how detected cases (c) and the EWT (d) depend on the percentage of population tested for $4$ surveillance strategies: prioritizing critical mobility flows (blue) , prioritizing critical transportation stations (red), prioritizing the busiest transit hubs (black), and using a random distribution of resources (dashed grey line). In these panels, solid lines represent the average and the shadowed region constitute the 95\% confidence interval for the results obtained for each fraction of population tested across 100 simulations.}
    \label{fig:transport_links}
\end{figure*}


We further delineate the implications of timely control strategies following the epidemic alarm raised by mobility-informed surveillance campaigns. To do so, we consider that, when $n_{\text{alarm}}$ cases are detected, containment strategies are deployed, halving the reproduction number $\mathcal{R}_0$. Figure~\ref{fig:EWT}.f shows the trajectories of the epidemic after the activation of containment measures at the EWT ascertained through the surveillance strategies shown in Fig.~\ref{fig:EWT}.c. Significantly, in all instances, a marked decline in case numbers is observed, as indicated by the dashed lines in Fig.~\ref{fig:EWT}.b (that replicate the evolution of the average values shown in Fig.~\ref{fig:EWT}.f). Focusing on the differences among the scenarios presented in Fig.~\ref{fig:EWT}.c, we note that an anticipation of roughly 10 days in the EWT not only postpones the epidemic peak but also substantially curtails peak incidence. A more granular analysis of the effects of the containment measures implemented at the EWT is presented in Figure S4 in the Supplementary Material, which showcases a significant reduction in the size of the epidemic peak and the attack rate when focusing the tests on the most vulnerable flows, as well as a temporal shift of the epidemic peak.

\subsection{Mobility-informed surveillance at Transit Stations}\label{sec:transport}


So far, mobility-informed strategies have demonstrated effectiveness in early epidemic detection and in palliating their long-term burden on the population. However, the granular application of testing based on the anticipated knowledge of citizen's residences and workplaces is impractical during an actual epidemic due to privacy and logistical constraints. Consequently, other control strategies leveraging critical mobility data obtained from Mixing matrix ${\bf M}$, while still actionable, are necessary. To this aim, here we seek for alternative targeted strategies based on interventions at the level of transportation systems.


In Bogotá, the \textit{Transmilenio} bus rapid transit system accounted for 38\% of urban travel in 2022~\cite{2022transmilenio}. Its extensive network, shown in Fig.~\ref{fig:transport_links}.a, vertebrates the daily flows through 12 routes that feature 144 stations with varied geographic distribution. The system's electronic card check-ins and check-outs offer a digital trace of users' travel patterns. This feature, along with the massive use of the Transmilenio mobility network, facilitates the implementation of mobility-informed policies by allocating tests at strategic transport hubs. 


To identify those stations where resources should be prioritized, we need to map the critical flows identified in the mobility network, connecting two geographical patches, with critical journeys between two stations in the transportation network. To do so, we assume that the {\em Transmilenio} network is used by those individuals whose residence and workplace are located within a radius of $800$ meters of distance from their respective nearest stations, which are considered to be the origin and destination of the trip in the transportation network. Overall, the aggregation process, sketched in Fig.~\ref{fig:transport_links}.b and further explained in the Methods section, preserves $15\%$ of the human movements recorded in the mobility network.

Figures~\ref{fig:transport_links}.c-d present a comparison between allocating tests on the critical subgroups of individuals identified from either the mobility or the transportation networks. To this aim we show how detected cases (Fig.~\ref{fig:transport_links}.c) and EWT (Fig.~\ref{fig:transport_links}.d) depend on the number of links chosen, which is represented as the percentage of population they contain. The latter choice provides a fair comparison between the mobility and transport formalisms as they involve different amounts of people per link (for context, Figure S5 in the Supplementary Material shows the relation between the number of chosen mobility links considered until now and the population assigned to these trips). In both cases, detections (EWT) decrease (increases) as tests are more evenly distributed across the population, showing that targeted surveillance campaigns on vulnerable subgroups of individuals provides a better early assessment of an epidemic crisis. Note that the EWT yielded by the transportation network is typically longer than that observed for surveillance strategies designed on the mobility network while the number of detections through transit hubs are also systematically smaller than those attained analyzing critical mobility links. These are expected results, since different mobility flows, corresponding with subgroups of diverse exposure levels, can be aggregated into single transportation routes. Yet suboptimal, targeting specific users of the transportation network is a feasible surveillance strategy. More importantly, this strategy enables a much earlier anticipation of epidemic outbreaks than other policies not leveraging the information obtained from the mixing matrix ${\bf M}$, such as the indiscriminate city-wide distribution of resources (random) or prioritizing the busiest mobility flows in the transport network (population), both also reported also in Figures~\ref{fig:transport_links}.c-d.

\section*{Discussion}\label{sec:discussion}


In this work, we have explored the intersection of urban mobility and epidemic surveillance, harnessing the capacity of metapopulation frameworks to enrich epidemic models with a, often overlooked, layer of behavioral complexity. Our findings contribute to the emerging field of digital epidemiology~\cite{salathe2018digital}, addressing the challenge of incorporating human behavioral data to construct realistic human mixing models that improves our preparedness and response to epidemic scenarios~\cite{agenda}.



We have shown that urban mobility flows and demographic data can be combined into a Mixing matrix that captures the interactions of residents in urban environments driven by their daily recurrent mobility patterns. Our results underscore that the dominant eigenvector of this matrix has the potential to streamline testing and quarantine efforts by identifying critical pathways conducive to pathogen transmission. Thus, equipped with the particular mixing matrices of real cities, we have leveraged the knowledge of critical flows to design targeted mitigation measures (based on test and quarantine policies) and surveillance campaigns. In both scenarios, we have assumed limited testing resources, a common situation when facing sanitary crises in large urban environments, and studied how the allocation of testing resources to the most critical flows affects the outcomes of both processes.


For containment purposes, we have demonstrated that an optimal concentration of testing resources can significantly reduce the impact of an epidemic outbreak. However, the contrasting scenarios observed in cities like Bogotá and Miami illustrate that the optimal distribution of tests is contingent upon the specific socio-economic urban landscape. This insight is critical for urban planners and public health officials seeking to enhance the resilience of cities against future outbreaks.

In the realm of surveillance, we have shown that the concentration of testing resources in critical mobility flows results in a considerable anticipation of the epidemic wave, allowing for prompt action even with limited resources. Specifically, the concept of Early Warning Time (EWT) emerges as a key metric for assessing the timeliness of epidemic alerts, reinforcing that targeted testing of individuals within high-risk transmission flows can lead to more expedient interventions. We have also demonstrated that the proposed human behavior-informed surveillance campaigns are significant and actionable. By showing that strategic testing at public transport hubs — a more practical approach than widespread testing — can be an effective method of epidemic control, we provide a feasible blueprint for urban health policy design.

In conclusion, our research highlights the importance of strategically leveraging human behavior data to design efficient control policies when facing an epidemic crisis~\cite{bubar2021model}. While challenges in the implementation of such strategies persist, particularly regarding data privacy and ethical considerations, the advantages of enhancing epidemic intelligence through astute data analysis are clear. As globalization advances and urban populations continue to expand \cite{globalization}, the necessity for informed and multidisciplinary approaches such as the one presented here becomes increasingly vital, providing a shift from merely reacting to epidemic crisis to an integrative program that includes, among others, epidemiology and social sciences~\cite{agenda,outlook}.

\begin{acknowledgments}
P.V. and J.G.G. acknowledge financial support from the Departamento de Industria e Innovaci\'on del Gobierno de Arag\'on y Fondo Social Europeo (FENOL group grant E36-23R), and from Ministerio de Ciencia e Innovaci\'on through project PID2020-113582GB-I00/AEI/10.13039/501100011033. A.F.U aknowledges funding support from the Department of Industrial Engineering, at Universidad de los Andes, Colombia through the doctoral training program. A.A. and D.S.P acknowledge support from Spanish Ministerio de Ciencia e Innovaci\'on (PID2021-128005NB-C21), Generalitat de Catalunya (2021SGR-00633) and Universitat Rovira i Virgili (2023PFR-URV-00633), the European Union’s Horizon Europe Programme under the CREXDATA project. D.S.P. acknowledges the financial support of the Calouste Gulbenkian Foundation through the PONTE program and from Ministerio de Ciencia e Innovaci\'on through the Juan de La Cierva program. A.A. acknowledges the Joint Appointment Program at Pacific Northwest National Laboratory (PNNL). PNNL is a multi-program national laboratory operated for the U.S. Department of Energy (DOE) by Battelle Memorial Institute under Contract No. DE-AC05-76RL01830, grant agreement no. 101092749,  ICREA Academia, and the James S.\ McDonnell Foundation (Grant N.\ 220020325). FM aknowledges the financial support of the  School of Engineering at Universidad de los Andes.
\end{acknowledgments}

\section*{Methods}\label{sec:methods}

\subsection*{Data}
Our framework represents cities as metapopulations whose nodes correspond to residence areas and their links encodes the mobility patterns of the population. The construction of a metapopulation then requires demographic data, capturing how residents are distributed across the city, and mobility data encoding their commuting patterns. For both cities here analyzed, we can construct their associated metapopulation by using publicly available data.

Demographic data for the city of Bogot\'a, Colombia, come from the 2018 National Census of Population and Housing~\cite{BogotaCensus}, involving 7 129 506 residents. The public database provides household, residence, and individual data at the spatial resolution of \textit{city blocks} (\textit{manzanas}). Moreover, origin-destination data ${\bf T}$ for commuters is sourced from the 2018 Bogot\'a mobility survey~\cite{BogotaMovilidad} with the resolution of the \textit{Transport Analysis Zones} (\textit{Zonas de Análisis de Transporte} or ZATs). These zones are considerably larger in area than the \textit{city blocks}. To match both data sources, we use available geometric spatial data to aggregate the populations of the \textit{city blocks} within each ZAT and use the latter as the patches of our metapopulation model, resulting in a mobility network with 826 nodes and 54890 links.

We also analyze the mixing matrix corresponding to the metropolitan statistical area (also known as core based statistical areas or CBSAs) spanning the cities of Miami, Fort Lauderdale and Pompano Beach, referred to as Miami for convenience throughout the manuscript. The 2010 census in the US accounts for $5590269$ people living in these cities and provides the distribution of population~\cite{UScensus} across \textit{census blocks}. The census also provides mobility data on the Longitudinal Employer-Household Dynamics~\cite{UScensusLEHD} (LEHD) database. Specifically, daily mobility patterns of the population can be extracted from LEHD Origin-Destination Employment Statistics~\cite{UScomm} (LODES), which provides commuting data over the entire country, also at the level of \textit{census blocks}. As census blocks represent very small geographical areas, we aggregate both data sets to ZIP Code Tabulation Areas (ZCTAs) resolution, thus being able to provide a more coarse-grained description of the locations on which resources should be prioritized. The aggregation process gives rise to a metapopulation with 186 nodes (ZCTAs) and 31791 links that connect them.

For both cities, the constructed metapopulations are fully characterized by the number of residents in each patch $i$, $n_i$ and the mobility flows $n_{ij}$ across patches, which can be constructed from the elements of the origin-destination matrix ${\bf T}$ as: 
\begin{equation*}
    n_{ij} = n_i\frac{T_{ij}}{\sum_{k}T_{ik}}\;.
\end{equation*}

\subsection*{Agent-based simulations}
All the epidemiological curves shown in the manuscript are computed through agent-based simulations combining mobility data $n_{ij}$ and the epidemiological processes driving the evolution of epidemic outbreaks. These simulations allow tracking the epidemiological state of each individual of the population according to a SEIRQD (Susceptible-Exposed-Infected-Recovered-Quarantined-Detected) dynamics. 

We use a discrete-time approach considering that each time step in the simulations corresponds to a day. For each time step, different processes are simulated: 

\begin{itemize}
\color{black}
    \item First, we distribute the population according to the mobility flows $n_{ij}$ recorded in the metapopulation, giving rise to a new (temporary) spatial distribution in which each patch $i$ is effectively populated by $n_i^{\textrm{eff}}= \sum_j n_{ji}$ agents. 
    
    \item Then, day contacts are simulated at each patch. These contacts aim at capturing interactions occurring at workplaces, schools etc. We assume that all individuals make the same number of contacts, which are proportional to the effective population density of the patch in which they are located. Therefore, an individual located at patch $i$ makes $z_D f_i$ interactions with randomly chosen individuals within their temporary patch, where $f_i=n_i^{\textrm{eff}}/a_i$ captures the effective population density at patch $i$, being $a_i$ its area. Likewise, $z_D$ is a scaling factor to ensure that the average number of day contacts is $\langle k_D \rangle$, i.e. $z_D=\langle k_D\rangle/\sum_i n_i^{\textrm{eff}} f_i $.
    Throughout the manuscript, we assume $\langle k_D\rangle=8$ for all epidemic scenarios.
    To simulate contagions, we assume that a susceptible (S) individual becomes exposed (E) with a probability $\beta$ for each contact with an infected (I) person. 
    
    \item Then, we simulate contagion processes with household members. Those processes add a total number of $\langle k_{H} \rangle$ interactions to every susceptible individual, which are established with randomly chosen individuals within their residence patch. Throughout the manuscript, we assume $\langle k_H\rangle=3$ for all epidemic scenarios.
    
    \item The rest of compartments get updated according to their respective dynamics. Every exposed (E) individual has a probability $\eta$ of turning infected (I), and every infected (I) agent has a probability $\mu$ of becoming recovered (R).
    \item Finally, depending on the control strategy in question, a specific number of tests $n_{test}$ are randomly distributed over agents belonging to the selected links in the mobility network, $n_{ij}$, or in the transportation network $n_{IJ}$ in case of acting on public transport users. When enforcing quarantine, every positive test will set the state of the individual to a different compartment, quarantined (Q), isolating them from the rest of the population.
\end{itemize}

The parameters used for all the simulations and the deterministic equations that describe the dynamics of the system can be found in the Supplementary Table 1 and Appendix A.

\subsection*{Mixing matrix M}

As explained in the main text, the Mixing matrix ${\bf M}$ simplifies the multitude of mobility-driven contagion processes into a single, mathematically manageable entity, whose dimension corresponds to the number of different subgroups in the metapopulation under study. In particular, we focus on the Mixing matrix ${\bf M}$ when $p_d=1$, for it captures the essence of behavioral-driven surveillance in urban landscapes subjected to the baseline mobility scenario. 

To construct this matrix, we must take into account the two different types of contagion processes occurring at each time step in our model. First, we assume that individuals interact in their destination, being their number of contacts proportional to the effective population density inside that patch. Given a destination patch $i$, the effective population density $f_i$ is given by $f_i=n_i^{\textrm{eff}}/a_i$, where $n_i^{\textrm{eff}}$ represents the effective population of patch $i$, i.e. $n_i^{\textrm{eff}}=\sum_j n_{ji}$, and $a_i$ encodes the area of the patch $i$. In addition, we assume that individuals make $\langle k_H \rangle$ contacts with individuals from their household.

From these assumptions and as shown in Appendix B, the element $M^{ij}_{kl}$, encoding all the interactions of one individual whose residence (destination) is located in $i$ ($j$) with the subgroup of residents in $l$ travelling to $k$, is given by:
\begin{align}
\label{eq:mixing}
M^{ij}_{lk}&=(1 - p_d)^2 \frac{z^D f_l}{n_l^{\text{eff}}} n_{lk}\delta_{il} + \frac{\langle k_H\rangle}{n_l}n_{lk}\delta_{il}\nonumber \\
&+(1 - p_d) p_d\left(\frac{z^D f_k}{n_k^{\text{eff}}}n_{lk}\delta_{ik}+\frac{z^D f_l}{n_l^{\text{eff}}}n_{lk}\delta_{jl}\right) \nonumber\\
&+p_d^2 \frac{z^D f_k}{n_k^{\text{eff}}}n_{lk}\delta_{jk}
\;.\;\;\;\;\;\;\;\;\;\;\;\;\;\;\;\;\;\;\;\;\;\;\;\;\;\;\;\;\;\;\;\;\;\;\;\;\;\;\;\;\;\;\;\;\;\;\;\;\;\;\;\;\;
\end{align}
where $\delta_{il}$ denotes the Kronecker delta and $z^D$ ensures that an average number of $\langle k_D \rangle$ daily contacts are observed across all destinations, i.e. $z_D = \langle k_D\rangle /\sum_i f_i n_i^{\text{eff}}$. 

\subsection*{Mobility-informed testing strategies and the Mixing matrix $M$}

The mixing matrix governs the evolution of the spatial distribution of infected individuals throughout the city at the initial stages of epidemic outbreaks, as proven in the Supplementary Material. For a Susceptible-Infected-Recovered (SIR) dynamics, defining $\epsilon_{ij}$ as the fraction of population with residence in $i$ and destination in $j$ in the infected state, the former evolution reads:
\begin{equation*}
\frac{\mu}{\beta}\epsilon_{ij}(t)=\sum_{lk}{\bf M}^{ij}_{lk}\epsilon_{lk}(t-1)\;,
\end{equation*}
where $\beta$ and $\mu$ denote the transmission and recovery probabilities. Expressing the former equation in matrix form, the evolution of the epidemic state of the population is given by the following linear equation:
\begin{equation*}
\frac{\mu}{\beta}\vec{\epsilon}(t)={\bf M}\vec{\epsilon}(t-1)\;.
\end{equation*}
The former expression neglects all nonlinear terms involved in contagions across the metapopulation, as it assumes a finite but negligible fraction of infected individuals across the metapopulation at early stages of the outbreak, i.e. $\epsilon_{ij} (t)\ll 1$ $\forall$ $i,j$. This assumption is no longer valid when the epidemic prevalence is higher across the metapopulation; however, it allows us to envisage testing policies relying on the spectral policies of the Mixing matrix $\bf{M}$. In particular, its leading eigenvector is expected to capture the spatial distribution of infected agents at the onset of an outbreak $\vec{\epsilon}$, thus providing an early estimate of the most vulnerable population subgroups. As the latter correspond to individuals with different origins and destinations, we will refer to the elements of the vector as links $L$ (origin-destination pairs) of the metapopulation.

\subsection*{Spatial allocation of tests}
Mobility-informed policies relying on the Mixing matrix $\bf{M}$ are shaped by two parameters: $n_{\text{tests}}$, determining the amount of resources available for testing the population, and $L$, dictating how many subgroups of individuals (links) are prone to be tested. Specifically, for each time step (day) in the agent-based simulations, $n_{\text{tests}}$ are distributed randomly among the pool of population composing the most critical $L$ links identified by analyzing the Mixing matrix $\bf{M}$. Specifically, following our analysis in the previous section, we assume that these links correspond to the $L$ largest entries of the leading eigenvector of matrix $\bf{M}$. This methodology allows us to evaluate whether concentrating resources on the most critical flows or spreading them out across the city constitutes an optimal strategy to face an emerging epidemic.

To compare our mobility-informed policies with other prioritization strategies, we assume that the spatial allocation of tests is determined by the fraction of the population included in the pool of individuals prone to be tested, rather than by the number of links $L$ targeted in the metapopulation. This assumption guarantees a fair comparison for testing strategies designed on different mobility networks. Beyond mobility-informed policies that act on the links of the mobilility network, three different strategies are implemented in Figure 5: {\em Random}, representing a random allocation of resources across the metapopulation, {\em Population}, prioritizing tests on the most crowded subgroups (links) in the mobility network and {\em Transportation}, prioritizing the most vulnerable subgroups found in the public transport network as explained in the next subsection.

\subsection*{Testing in the public transport network}

While quite useful for surveillance purposes and the long-term control of epidemic outbreaks, mobility-informed policies that rely on the Mixing matrix $\bf{M}$ are not feasible due to practical difficulties in identifying specific subgroups of the population. As explained in the main text, we can take advantage of the information provided by the Mixing matrix $\bf{M}$ and implement a more plausible testing policy by allocating strategically test in the transport stations, based on the passengers' entrance and exit. For this purpose, we need to project the information obtained from $\bf{M}$ to estimate the likelihood of a specific subgroup of users in the transportation network being infected. We focus our analysis on the city of Bogot\'a, for which we have extremely fine-grained demographic information from census data, the daily mobility patterns of the population from public surveys and the spatial distribution of the \textit{Transmilenio} network, the Bus Rapid Transit network of the city. 

To aggregate the mobility flows into the transport network, we use geographical location data of every Transmilenio station and assume people often use the closest stations to their origin and destination patches, in case they are located within a radius of 800 meters from their associated city blocks. Therefore, we assign a set of weights $W_{iI}$ to every patch $i$, representing the fraction of residents there using a nearby station $I$ to travel to their usual destination. Oftentimes, every person from a patch $i$ has the same nearest transport station $I$, i.e. $W_{iI}=1$. Nonetheless, there are also scenarios, as the one visualized in Fig.~\ref{fig:transport_links}b, for which patches are surrounded by different nearby stations, which are chosen disparately by the residents across the city blocks of those areas. In that case, weights are assigned as the proportion of residents who choose each of the stations.

Assuming that there is no correlation between the origin and destination stations, the number of individuals travelling from node $i$ to $j$ using the transportation line from station $I$ to $J$ is:
\begin{equation*}
    n_{iI}^{jJ} = n_{ij} W_{iI} W_{jJ}\;,
\end{equation*}
which works for the number of infected agents in the same way:

\begin{equation*}
    I_{iI}^{jJ} = I_{ij} W_{iI} W_{jJ}\;.
\end{equation*}

Adding the populations of all the nodes that use a particular pair of stations gives us the projection of the mobility network onto the transport network. Therefore, we can compute the fraction of the population infected at the onset of the outbreak and using the transportation route connecting stations $I$ and $J$, $\epsilon_{IJ}$, as

\begin{equation}
        \varepsilon_{IJ} = \dfrac{I_{IJ}}{n_{IJ}} = \dfrac{\sum_{ij} I_{jJ}^{iI}}{\sum_{ij}n_{jJ}^{iI}} = \dfrac{ \sum_{ij} W_{iI} W_{jJ} n_{ij} \varepsilon_{ij} }{\sum_{ij} W_{iI} W_{jJ} n_{ij}}\;.
\end{equation}

Note that the proposed projection loses some accuracy in the localization of critical mobility flows because it can group people from different links in the same transport line. Nonetheless, as argued above, it represents a realistic approach that is very useful in diagnosing and tackling the progression of an emerging epidemic wave. The results shown in the main text using this criterion were obtained in the same way as previously described: ranking the transport lines according to their risk $\varepsilon_{IJ}$ and targeting the most vulnerable ones when distributing the tests.

\section*{References}
\bibliography{aipsamp}

\renewcommand{\figurename}{Figure}
\renewcommand{\thefigure}{S\arabic{figure}}
\setcounter{figure}{0}    

\newpage

\section*{Supplementary Material}

\subsection*{Appendix A. Markovian formulation of the Movement-Interaction-Return (MIR) model}
\title[Supplementary Material: Human behavior-driven epidemic surveillance in urban landscapes]{Human behavior-driven epidemic surveillance in urban landscapes}
The mechanistic stochastic simulations of the Movement-Interaction-Return (MIR) model used to obtain the epidemic curves throughout the manuscript can be casted into a set of discrete-time deterministic equations. Restricting ourselves to a Susceptible-Infected-Recovered (SIR) dynamics, the state of the system at time $t$ can be described by a vector $\rho_{ij}(t)$ of size $N^2$ (being $N$ the number of subpopulations in the system), where every component represents the probability of an individual with origin $i$ and destination $j$ being infected at time $t$, and a vector $r_{ij}(t)$ that describes the probability of said individual being recovered. The time evolution of the system is characterized by the equations
\begin{align}\label{eq:MIR1}
&\rho_{ij}(t+1) = (1-\mu)\rho_{ij}(t) + \left(1 - \rho_{ij}(t) - r_{ij}(t)\right)\Pi_{ij}(t) &\forall \{i,j\}\in L \;,\\
&r_{ij}(t+1) = r_{ij}(t) + \mu\rho_{ij}(t) &\forall \{i,j\}\in L \;.
\end{align}
The first term in the r.h.s of the first equation represents the probability that the infected population does not recover, where $\mu$ is the probability an infectious agent recovering each day, stepping into the recovered compartment. Conversely, the second term encapsulates the new infections of susceptible agents. The probability that an individual with origin $i$ and destination $j$ contracts the disease, $\Pi_{ij}$, is calculated as follows:
\begin{multline}
\label{eq:MIR2}
\Pi_{ij}(t) = (1 - p_d)\left[ P_i^D(t) + (1 - P_i^D(t))P_i^N(t)\right] + \\p_d\left[ P_j^D(t) + (1 - P_j^D(t))P_i^N(t)\right]\;.
\end{multline}

The first term corresponds to contagions occurring for those individuals staying in their residence area, with probability $1-p_d$, whereas the second one involves those contagions affecting individuals moving with probability $p_d$ to their destination $j$. The equation assumes that a susceptible individual staying in patch $i$ (moving to patch $j$) during the day can contract the disease through their interactions made there with probability $P_i^D(t)$ ($P_j^D(t)$).  The equation also considers a second probability of infection after they return to their residence $P_i^N(t)$ whether or not they decided to travel. This term reflects the probability of contracting a disease following contacts with members from the same household. The probability of contagion in a given patch $i$ during the day $t$ is
\begin{equation}
\label{MIR3}
P_i^D(t) = 1 - \left( 1 - \beta\dfrac{I_i^{eff}(t) }{n_i^{eff}}\right)^{z^D f_i}\;.
\end{equation}
This equation assumes all agents inside the patch have the same connectivity and their contacts are randomly established. Here, $\beta$ represents the \textit{infectivity} of the disease or the probability that an agent contracts the disease when interacting with an infected one, $I_i^{eff}$ is the effective number of infectious agents in the patch and $n_i^{eff}$ its effective population. The number of interactions for each agent is given by $z^D f_i$, where $f_i$ is the effective population density in patch $i$, i.e. $f_i=n_i^{eff}/a_i$, being $a_i$ the area of patch $i$. Besides, $z_D$ is a scaling factor to set the average number of contacts to $\langle k_{D}\rangle $, i.e. $z_D = \langle k_{D}\rangle /\sum_i f_i n_i^{eff}$. The effective number of individuals and infected population in each patch $i$ at time $t$ read:
\begin{gather}
    n_i^{eff} = \sum_{j=1}^N \left(p_d n_{ji} + \left( 1 - p_d \right) n_{ij} \right)\;,\\
    I_i^{eff}(t) = \sum_{j=1}^N \left(p_d n_{ji} \rho_{ji}(t) + \left( 1 - p_d \right) n_{ij} \rho_{ij}(t) \right)\;,
\end{gather}

Analogously, we can write the probability that a susceptible individual from patch $i$ contracts the disease when interacting with members from their household at time $t$, $P_i^N (t)$, as
\begin{equation}
\label{MIR4}
P_i^N(t) = 1 - \left( 1 - \beta \frac{\displaystyle\sum_{j=1}^N n_{ij}\rho_{ij}(t)}{\displaystyle\sum_{j=1}^N n_{ij}} \right)^{\langle k_H \rangle}\;,
\end{equation}
where $\langle k_H \rangle$ represents the average number of contacts with household members.

\subsection*{Appendix B. The mixing matrix ${\bf M}$ as driver of epidemic outbreaks}

Here we show that the mixing matrix ${\bf M}$ governs the evolution of the spatial distribution of infected individuals at early stages of an epidemic outbreak. To do so, let us assume that, at the beginning of an outbreak, infected individuals represents a negligible fraction of the total population in each subgroup found in the metapopulation. Therefore, we assume $\rho_{ij}(t) \equiv \varepsilon_{ij}(t) \ll 1 \quad \forall i,j$. The former assumption allows neglecting $O(\epsilon^2)$ terms and linearizing Eqs.(\ref{MIR3}-\ref{MIR4}), which now read:
\begin{eqnarray*}
P_i^D(t) &\simeq& p_d \beta \dfrac{z^D f_i}{n_i^{eff}}\sum_{j=1}^N n_{ji}\varepsilon_{ji} + \beta (1 - p_d) \dfrac{z^D f_i}{n_i^{eff}} \sum_{j=1}^N n_{ij} \varepsilon_{ij}\;, \\
P_i^N(t) &\simeq& \beta \dfrac{\langle k_H\rangle}{n_i} \sum_{j=1}^N n_{ij}\varepsilon_{ij}\;.
\end{eqnarray*}
Plugging this expression into Eq.~(\ref{eq:MIR2} and neglecting nonlinear terms, we obtain
\begin{equation}
\Pi_{ij}(\vec{\epsilon}) = \beta \sum_{l,k=1}^N M_{lk}^{ij} \varepsilon_{lk}\;.\;\;\;\;\;\;\;\;\;\;\;\;\;\;\;\;\;\;\;\;\;
\label{pi_ij}
\end{equation}
where $M^{ij}_{lk}$ corresponds with the elements of the mixing matrix which are given by:
\begin{multline}
M^{ij}_{lk} \equiv \left.(1 - p_d) p_d \frac{z^D f_k}{n_k^{eff}}n_{lk}\delta_{ik} + (1 - p_d)^2 \frac{z^D f_l}{n_l^{eff}} n_{lk}\delta_{il} \right.\\
 \left.+p_d^2 \frac{z^D f_k}{n_k^{eff}}n_{lk}\delta_{jk} +p_d(1 - p_d)\frac{z^D f_l}{n_l^{eff}}n_{lk}\delta_{jl} + \frac{z^N \sigma_l}{n_l}n_{lk}\delta_{il} \right.\;\;\;\;\;\;\;\;\;\;\;\;\;\;\;\;\;\;\;
\end{multline}
Finally, introducing the former expression into Eq.~\ref{eq:MIR1}, neglecting the recovered population, i.e. $r_{ij}=0$, and keeping just the linear terms yields:
\begin{equation*}
\frac{\mu}{\beta}\vec{\epsilon}(t)={\bf M}\vec{\epsilon}(t-1)\;,
\end{equation*}
proving that the mixing matrix ${\bf M}$ governs the onset of epidemic outbreaks. 
\newpage 
\newpage

\begin{table*}[t!]
\centering
\caption{Values of the fixed parameters used for all the simulations. The infectivity $\beta$ depends on the ${\cal{R}}_0$ to cover different situations, and is calculated from the largest eigenvalue of the mixing matrix ${\bf M}$.\\}
\label{tab:my-table}

\setlength{\tabcolsep}{15pt} 

\begin{tabular}{ccc}
\hline
\hline
Symbol                & Description                     & Value             \\
\hline
$\beta$               & Infectivity per contact         &${\cal{R}}_0 \mu / \Lambda_{max}({\bf M})$\\ 
$\eta$                & Latency probability                     & 1     \\
$\mu$                 & Recovery probability                 & 0.2   \\
$\langle k_D \rangle$ & Average daily contacts          & 8                 \\ 
$\langle k_H \rangle$ & Average home contacts           & 3                 \\
$p_d$                 & Probability of travel           & 1                 \\
\hline \hline
\end{tabular}
\end{table*}

\begin{figure*}[t]
    \centering
    \includegraphics[width=0.7\linewidth]{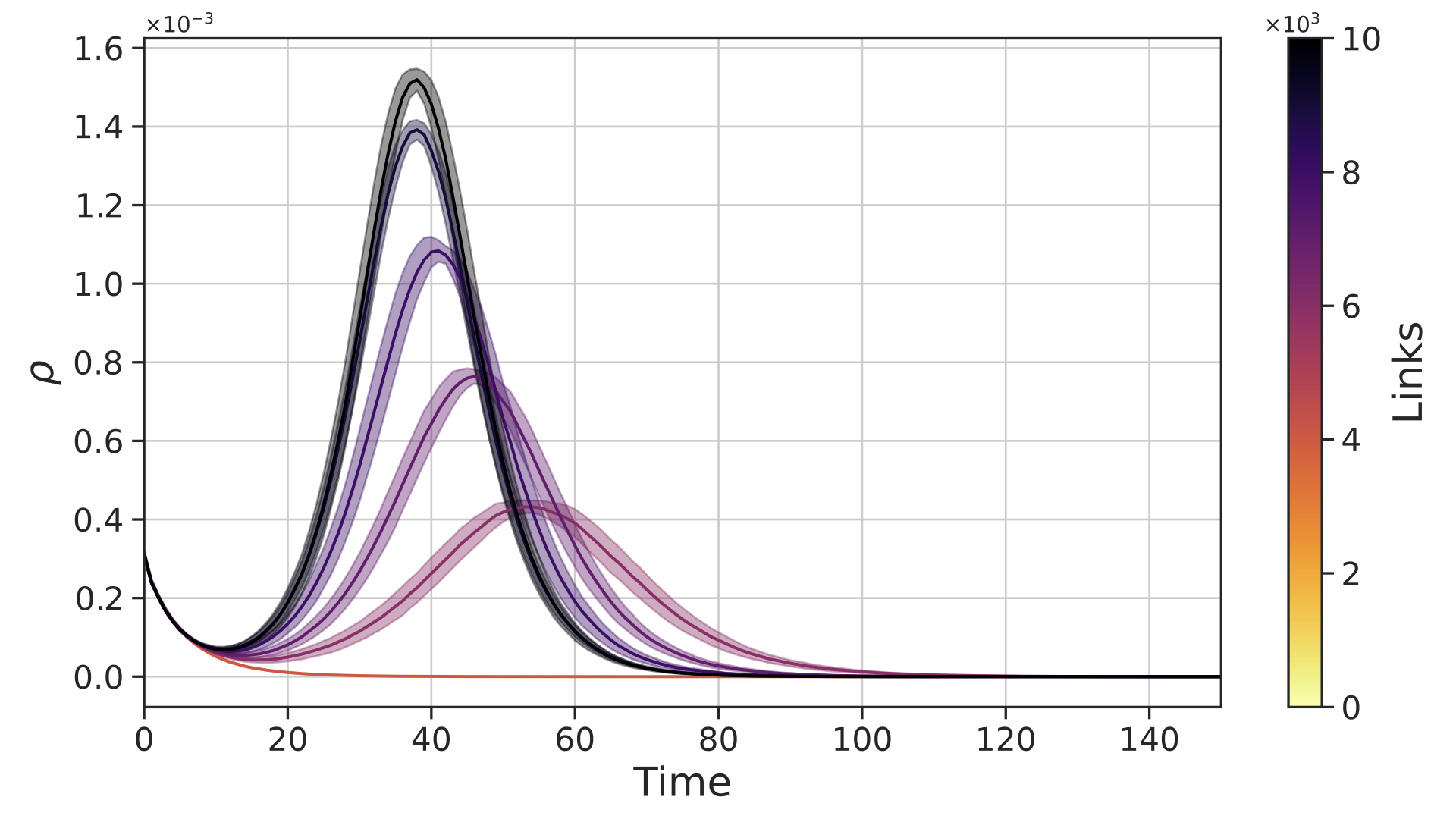}\caption{{\bf Epidemic curves under different targeted testing strategies in Miami}. Time evolution of the number of infected individuals in Miami when distributing $n_{tests}=2\times 10^5$ tests on the $L$ most critical human flows (color code), enforcing confinement to the agents who test positive. The basic reproduction number of the pathogen giving rise to the outbreak is ${\cal R}_0=4$. The shaded areas represent the 95\% confidence interval over 25 simulations.}
    \label{fig:extended1}
\end{figure*}

\begin{figure*}[t]
    \centering
    \includegraphics[width=0.75\linewidth]{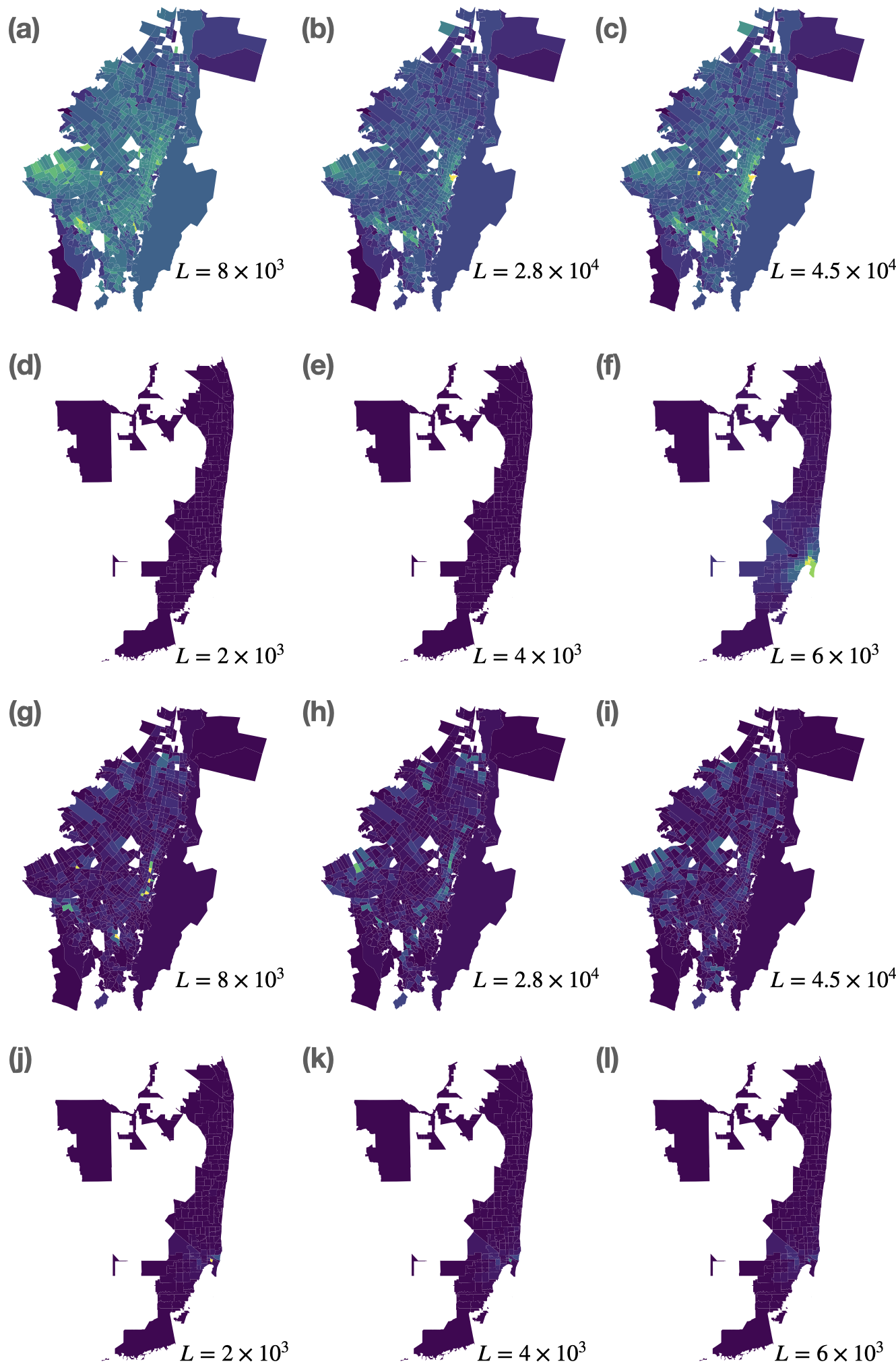}\caption{{\bf Spatial visualization of resource distribution and effects.} (a-c) Distribution of cases across Bogotá when allocating $2\times 10^5$ daily tests over $L = 8\times10^3$ , $2.8\times 10^4$ and $4.5\times 10^4$ links, respectively, and (d-f) across Miami when allocating $2\times 10^5$ daily tests over $L=2\times10^3$, $4\times10^3$ and $6\times10^3$, with ${\cal{R}}_0 = 4$ for all scenarios. (g-i) shows the spatial distribution of the tests according to the destinations of the chosen mobility flows in Bogot\'a and (j-l) Miami for the same amounts of selected flows as before. In all the maps, brighter colors correspond with higher values of the represented quantities.}
    \label{fig:extended2}
\end{figure*}

\begin{figure*}[t]
    \centering
    \includegraphics[width=0.95\linewidth]{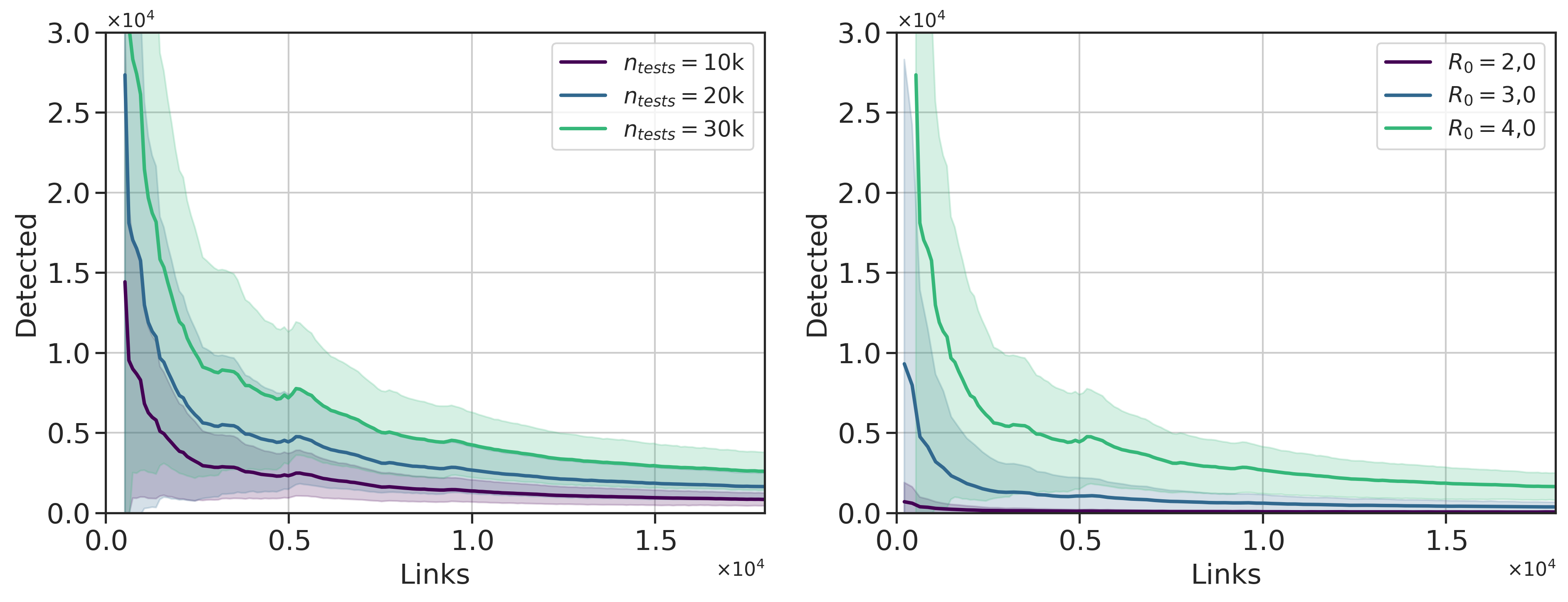}
    \caption{{\bf Effectiveness of targeted testing}. The total number of cases detected during the first 30 days of the epidemic wave plotted against the degree of test concentration for (a) various numbers of tests available, $n_{\text{tests}}$, with ${\cal{R}}_0 = 4$ and for (b) different basic reproduction numbers, with $n_{\text{tests}}=20\times 10^3$. The shadowed areas represent the 95\% deviation of the results from 100 simulations.}
    \label{fig:extended3}
\end{figure*}

\begin{figure*}[t]
    \centering
    \includegraphics[width=0.95\linewidth]{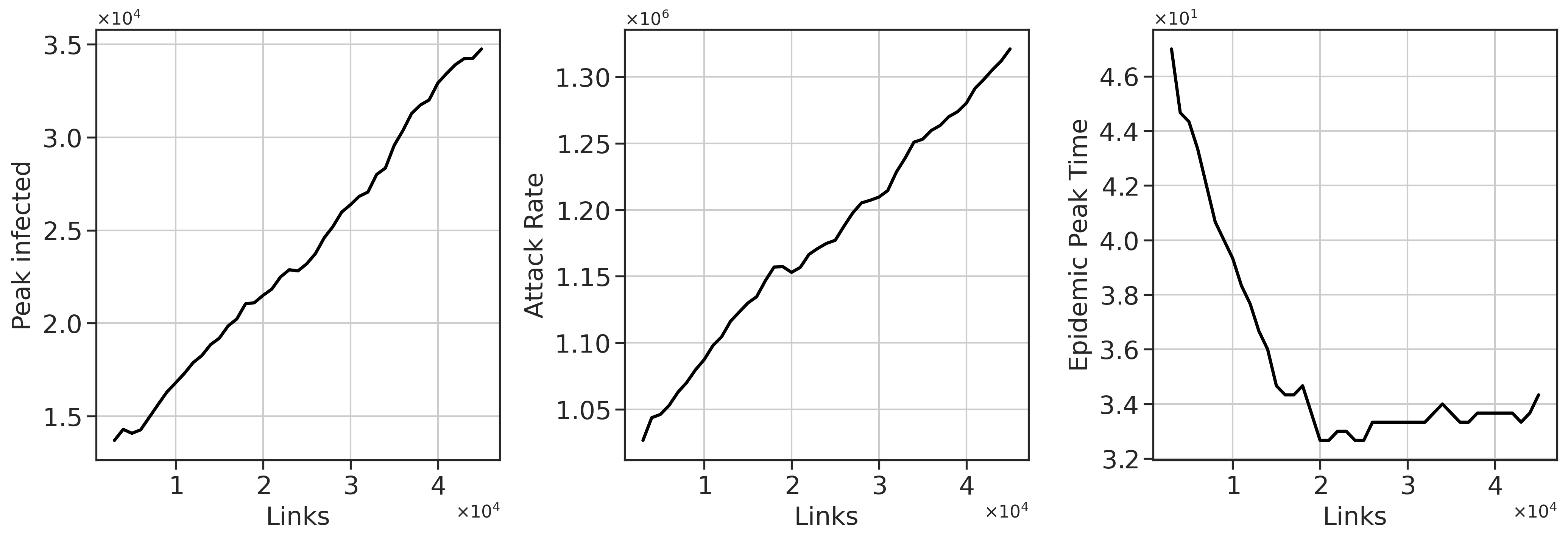}
    \caption{{\bf Relationship between different epidemic metrics and distribution of tests}. (a) Peak of infected individuals, (b) attack rate and (c) epidemic peak time of the mitigated epidemic curves (equivalent to the ones showed in Fig.~\ref{fig:EWT}f) as a function of the conncetration of links $L$ in the mobility network of Bogot\'a. In these outbreaks, ${\cal{R}}_0$ is reduced from $4$ to $2$ after reaching the epidemic warning time through containment measures.}
    \label{fig:extended4}
\end{figure*}

\begin{figure*}[h!]
\centering
\includegraphics[width=.5\linewidth]{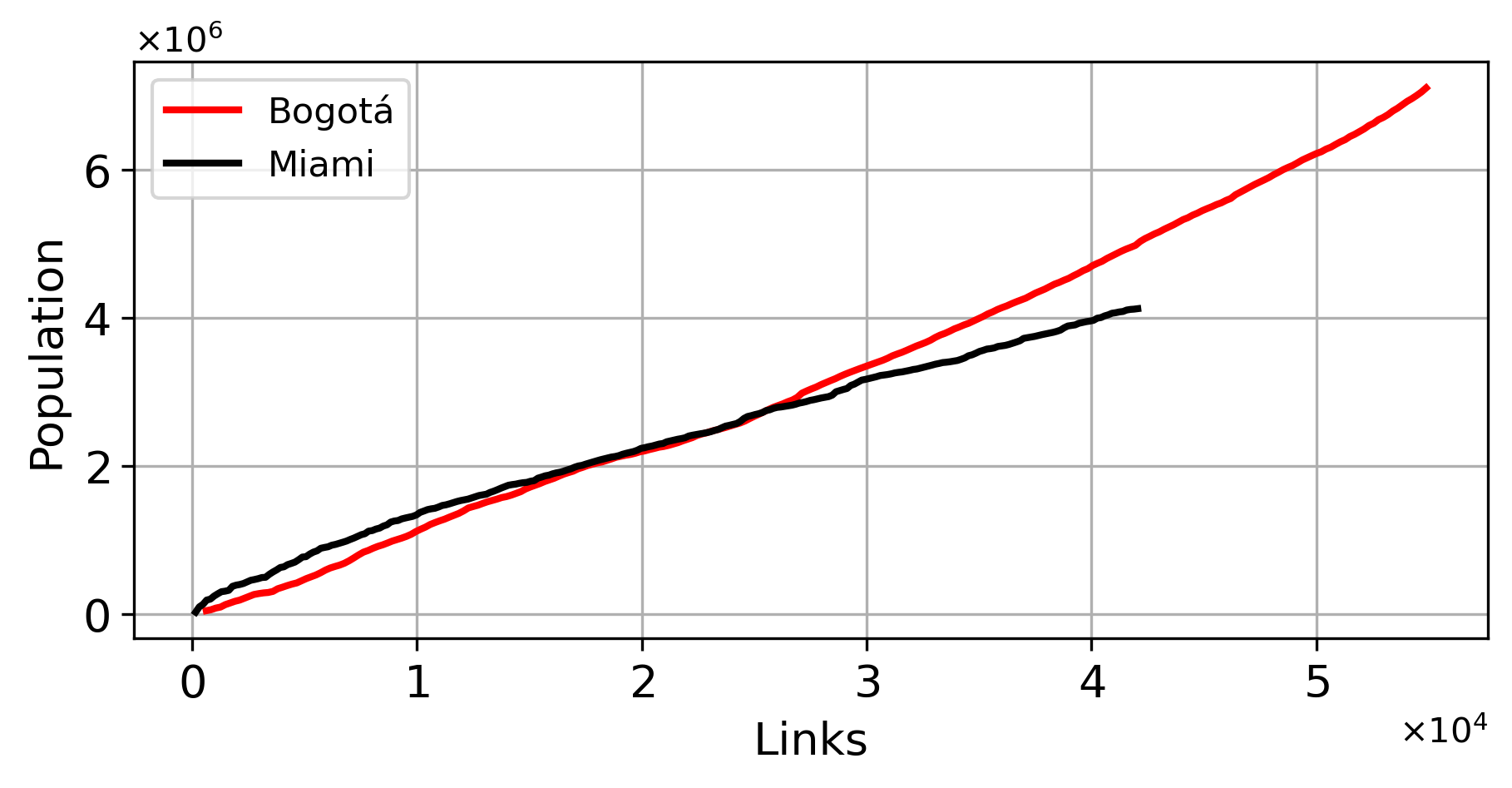}
\caption{{\bf Population distribution}. The cumulative population that belong to the chosen mobility flows plotted against the number of said flows, for the cities of Bogot\'a and Miami. The links are considerably heterogeneous in size and this prevents a linear relationship between both variables.}
\label{fig:nFig1}
\end{figure*}

\end{document}